\RequirePackage{fix-cm}
\documentclass{svjour3}                     % onecolumn (standard format)
\smartqed  % flush right qed marks, e.g. at end of proof
\usepackage{graphicx}% Include figure files
\usepackage{dcolumn}% Align table columns on decimal point
\usepackage{bm}% bold math
\usepackage{physics}
\usepackage{color}
\usepackage{hyperref}% add hypertext capabilities
\usepackage{amsmath}
\usepackage{amssymb}
 \usepackage{newtxtext,newtxmath} % {mathptmx} % 'mathptmx' is ancient  
 \usepackage{tabularx,caption,makecell}
 \usepackage[table, svgnames, dvipsnames]{xcolor}
 \usepackage[square,numbers]{natbib}
\usepackage[utf8]{inputenc}
\newcommand{\UFSC}{Departamento de Física, Universidade Federal de Santa Catarina,
88040-900, Florianópolis, SC, Brazil}

\newcommand{\IIP}{International Institute of Physics, Federal University of Rio Grande do Norte, 
59078-970, Natal, RN, Brazil}

\newcommand{\UFAL}{Grupo de F\'isica da Mat\'eria Condensada, N\'ucleo de Ci\^encias Exatas - NCEx, Campus Arapiraca, Universidade Federal de Alagoas, 57309-005, Arapiraca, AL, Brazil}
\begin{document}

\title{Automated Machine Learning can Classify Bound Entangled States with Tomograms%\thanks{Grants or other notes
}


\author{Caio B. D. Goes $^{1}$        \and
        Askery Canabarro $^{2,3}$ \and
        Eduardo I. Duzzioni $^{1}$ \and
        Thiago O. Maciel $^{1}$%etc.
}


\institute{Caio B. D. Goes\\
              \email{caio.boccato@posgrad.ufsc.br}           %  \\
           \and
      \\    $^{1}$  \UFSC
      \\   $^{2}$ \IIP
      \\   $^{3}$ \UFAL
}

\date{Received: date / Accepted: date}

\maketitle

\begin{abstract}
For quantum systems with total dimension greater than six, the positive partial transposition (PPT) criterion is necessary but not sufficient to decide the non-separability of quantum states. Here, we present an Automated Machine Learning approach to classify random states of two qutrits as separable or entangled even when the PPT criterion fails. We successfully applied our framework using enough data to perform a quantum state tomography and without any direct measurement of its entanglement. In addition, we could also estimate the Generalized Robustness of Entanglement with regression techniques and use it to validate our classifiers.
\end{abstract}

\section{Introduction}\label{sec:introduction}

Entanglement is critical for real-world Quantum Information protocols like Quantum Computation \cite{shor1994,vandersypen2001} and Unconditionally Secure Quantum Cryptography  \cite{Ursin2007,Vaziri2002,Grblacher2006}. In this context, non-separable systems with local dimensions with more than two degrees of freedom are especially important due to its tolerance to noise \cite{Grblacher2006}.

For quantum systems with total dimension up to six (\emph{e.g.}, two qubits or one qubit and one qutrit), the so-called Peres-Horodecki's \emph{Positive Partial Transposition} (PPT) criterion \cite{Peres1996,Horodecki1996} is necessary and sufficient. For larger systems, the criterion is sufficient but not necessary and we have entangled states with positive partial transposition \cite{Horodecki1998}. We can find several examples of PPT entangled states (PPTES), also known as bound entangled states, in the literature \cite{Horodecki1998,Horodecki1999,Bennett1999,Breuer2006,Sents2016,Shor2003}. 

Although we can decide whether a quantum state has \emph{Negative Partial Transposition} (NPT) or PPT within a polynomial number of operations, there is no operational necessary and sufficient criterion for the separability problem in general, moreover, Gurvits proved this decision problem to be NP-HARD \cite{Gurvits2003,Gurvits2003a}. 

In case we are willing to pay the computational effort, we can choose an \emph{Entanglement Witness} (EW) \cite{Horodecki1996} for entanglement detection. In terms of experimental setups, we can implement EWs when we have partial information about the state (\emph{cf.} the references in \cite{horodecki2009}), but they are not attractive, since each state has its own optimal witness.

It is not difficult to imagine real-world scenarios in which we will need to decide whether our quantum states are entangled or not in a very short time window. In spite of having numerous schemes to infer entanglement with partial information \cite{guhne2002,cavalcanti2006,maciel2009,lima2010}, they usually rely on EWs, especially when we have PPTES. In such time-sensitive scenarios, spending hours to compute optimal entanglement witnesses (OEW) is a luxury, so here we advocate the use of \emph{Machine Learning} (ML) techniques to characterize entanglement in an efficient post-processing manner, keeping the data gathering as simple as possible. 

ML techniques have recently become attractive tools to explore quantum mechanics in different problems \cite{Dunjko2018}, for instance, there are successful applications in problems like quantum tomography \cite{Kieferov2017,Torlai2018}, quantum error correction \cite{Torlai2017,Krastanov2017}, quantum control \cite{August2017}, assessment of NPT entanglement through Bell inequalities \cite{Gao2018,Yang2019,Canabarro2019}, assessment of separability aided by convex hull approximation \cite{Lu2018}, and quantum phase transitions \cite{Canabarro2019b}. In a nutshell, ML is used to build complex models capable of making predictions about difficult problems \cite{aaron}, such as the separability problem described above. 

Differently from \cite{Gao2018,Yang2019}, here we investigate how ML tools deal with harder instances of the classification of entanglement between two qutrits. More specifically, we do not consider a particular family of states with a small number of coefficients, we have a list of randomly drawn states classified as separable (SEP), entangled with positive partial transposition (PPTES), and entangled with negative partial transposition (NPT). 

With this list, we chose a scenario in which we possess enough data to perform a quantum tomography of a state $\rho$ via linear system inversion \cite{nielsen2000}, but no direct measurements of its entanglement. In \cite{Lu2018}, Lu \emph{et al.} used tomograms and a "distance" to the separable set using linear programming and convex hull approximation, which can be understood as an entanglement measure like the \emph{Robustness of Entanglement} \cite{vidal1999}. Using such direct measure of entanglement as a feature to train the machine might mask the real hardness of the problem: in principle, one could spend an enormous amount of resources to compute this measure, say $\alpha$, and $\alpha$ alone could give us the answer whether $\rho$ is entangled or not, so the final objective of the trained machine would be dispensable. With that said, we chose to test a harder venue using just tomograms, since they are not closely related to the object in question.

% , the final objective of the machine will be dispensable, since only with the measure that is usable as input is it possible to estimate entanglement of the state}. 
% which encodes the Hilbert space geometry as a feature in the machine's training to classify states of two qubits and two qutrits among separable and entangled. The algorithm proposed by the authors becomes very computationally costly since it will be necessary to provide two inputs, the tomagraphy and the entanglement measure, both costly.

The idea here was first assess the possibility of using a fewer number of features, but it was not possible for random states. Using all coefficients, we applied supervised ML techniques to infer a function that best maps the tomograms to our labels (SEP/PPTES/NPT). In spite of including NPT states in the list, we are particularly interested in the hardest instance of the problem: where Peres-Horodecki criterion cannot decide between SEP/PPTES. For this task, we chose \emph{Automated Machine Learning} (autoML) schemes \cite{le2019scaling,NIPS2015_5872,jin2019auto,mendoza-automlbook18a}  and evaluated their performances in this decision problem.

The computational effort to construct this training dataset is herculean, but it pays off when, after the completion of a successfully model, we can decide if a state $\rho$ is SEP/PPTES and SEP/PPTES/NPT with high probability. In addition, we could also estimate the amount of entanglement in the state using autoML for the regression task of an entanglement quantifier, namely, the \emph{Generalized Robustness of Entanglement}  \cite{steiner2003}. 

This paper is organized as follows: in Sec. \ref{sec:dataset}, we thoroughly explain our methodology for constructing our dataset; in Sec. \ref{sec:automl}, we explain the automated machine learning and its advantages; in Sec. \ref{sec:results}, we discuss our results and, finally in Sec. \ref{sec:conclusion}, we conclude and give some future perspectives.

\section{\label{sec:dataset} Dataset}

\begin{figure}
\includegraphics[width=\columnwidth]{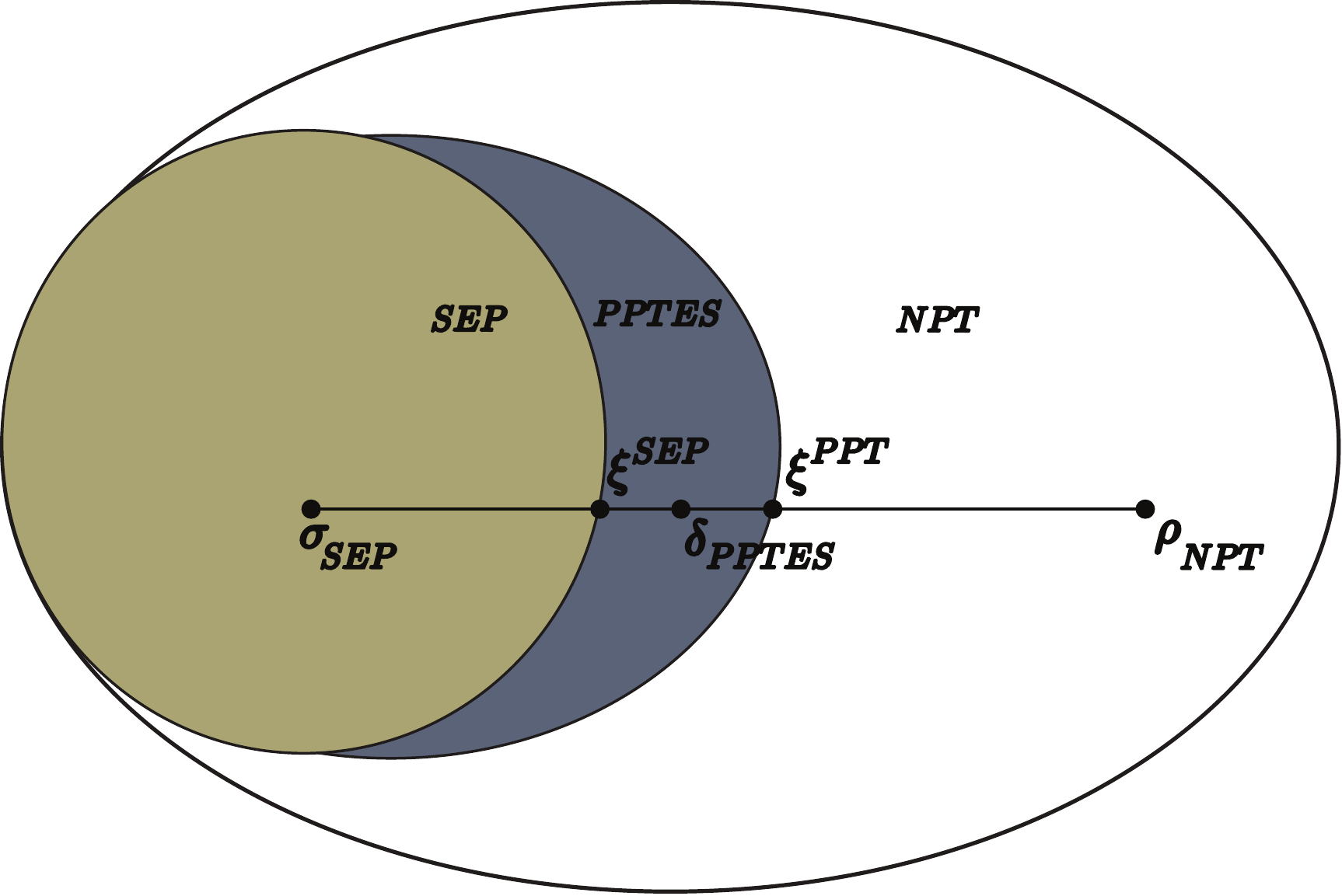}%
\caption{\label{fig:state-space}Schematic representation of state space with separable states (SEP) $\sigma_{SEP}$, entangled states with Positive Partial Transposition (PPTES) $\delta_{PPTES}$, and states with Negative Partial Transposition (NPT) $\rho_{NPT}$.  We also have edge states $\xi^{SEP}$ and $\xi^{PPT}$ lying in the frontier between SEP/PPTES and PPTES/NPT, respectively. }
\end{figure}

Choosing the right methodology to construct a training dataset is crucial to the quality of the results \cite{may2010} in any ML task. Sampling random mixed states of a given dimension can be non-trivial for different reasons: (i) we need to specify a probability measure on the set of all density operators of this dimension \cite{zyczkowski2011}; (ii) the volume of separable (SEP), entangled states with positive partial transposition (PPTES) and  entangled states with negative partial transposition (NPT) are not the same \cite{zyczkowski1998,zyczkowski2003,sommers2004}; (iii) there is no operational necessary and sufficient criterion for the separability problem in general.

Our methodological choices for these issues were the following: (i) we chose to sample our mixed states using the Hilbert-Schmidt measure \cite{braunstein1996,zyczkowski2001}; (ii) since SEP, PPTES and NPT sets have different volumes, we chose to use the same number of examples in each class; (iii) we first split between NPT/PPT using the Peres-Horodecki criterion \cite{Horodecki2001} and, within the PPT, we decided between PPTES/SEP using $\epsilon$\emph{-Optimal Entanglement Witnesses} ($\epsilon$-OEW) \cite{brandao2004a,Brandao2005} to compute the \emph{Generalized Robustness of Entanglement} (GR) \cite{steiner2003}. 

We opt for (ii) to balance our dataset, otherwise we would have a small number of SEP and PPTES in the sample \cite{zyczkowski1998}. Since the probability of randomly draw a PPTES is low, we used two strategies to sample in this set, explained in the next sections.

It is worth to note that this work differs from \cite{Gao2018,Yang2019} in several aspects. Beside the choice of using random states, we also used a different choice of non-separability tests when Peres-Horodecki criterion fails. We chose EWs instead of Doherty \emph{et al.} \cite{doherty2004} $k$-separability hierarchy so we could estimate the amount of entanglement in the state $\rho$ for further use in the regression test. We also had numerical evidence that using just the lower level ($k=2$) of the hierarchy lead to even lower probability of asserting PPTES of randomly drawn states, with no addition to the performance in Section \ref{sec:results}.

\subsection{Dataset with random states of two qutrits}

Suppose we draw a random state $ \rho \in \mathcal{D}  \subset \mathcal{L} (\mathbb{C} ^{3} \otimes \mathbb{C} ^{3}) $, where $ \mathcal{D} $ is the state space, and $\mathcal{L} $ is the set of linear operators in the Hilbert space $\mathcal{H} \simeq \mathbb{C}^{3} \otimes \mathbb{C}^{3}$.  

We can write $\rho$ as
\begin{equation}\label{eq-complete-rho}
\rho = \sum_{i=1}^{d^2-1} \theta_i E_i,
\end{equation}
where $E_i$ form a basis in Hilbert-Schmidt space and $\theta_i$ is the solution to the following linear system \cite{nielsen2000}:
\begin{equation}\label{eq-linear-inversion}
G^{-1}\vec{c} = \vec{\theta}, \quad G_{ij} = \trace{\left(E_i\,E_j \right)}, \quad c_i = \trace{\left(E_i\,\rho \right)}.
\end{equation}

We can choose any basis as long as it spans the Hilbert-Schmidt space. Here, we want to avoid measurements on entangled projectors, so we chose the $SU(3) \otimes SU(3)$ generators to simplify the measurement requirements in contrast, for example, to the $ SU (9) $ generators in \cite{Lu2018}.

The Generalized Robustness (GR) \cite{vidal1999,steiner2003} is an entanglement quantifier obtainable via  entanglement witnesses \cite{brandao2004a,Brandao2005}. GR is the minimal amount of mixing with any state needed to destroy the entanglement in $rho$, \emph{i.e.},

\begin{equation}\label{eq:gr}
GR(\rho) = \min_{\sigma , \in \, \mathcal{D}} \left( \min_{s \geq 0 \, \in \, \mathbb{R}} s: \xi^{SEP} \equiv \frac{\rho + s \sigma}{1 + s}  \in SEP \right),
\end{equation}

where $\xi^{SEP}$ belongs to the edge between separable and entangled states (\emph{cf.} Fig \ref{fig:state-space}). Our $\epsilon$-OEW for GR guarantees that $\rho$ is entangled iff $GR(\rho) > \epsilon$, here we chose $\epsilon = 10^{-5}$.

For our learning process, we encoded $\rho$ in the real vector $\vec{c} \in \, \mathbb{R}^{d^2-1}$ (\emph{cf.} Eq. (\ref{eq-linear-inversion})). For better adherence to real world applications, we chose tomographic measurements, guaranteeing we have all needed information of $\rho$ but not direct measurements of its entanglement.

We can summarize the algorithm to generate our dataset as follows:

\begin{enumerate}
	\item Draw a random density matrix $\rho$ according to the Hilbert-Schmidt measure;
	\item Compute $GR(\rho)$ using an $\epsilon$-OEW and obtain its edge state $\xi^{SEP}$;
	\item Check whether $\rho$ is a PPT state;
	\item If $\rho$ is PPT and $GR(\rho) > \epsilon$, we save $\rho$ in our dataset as PPTES;
	\item If $\rho$ is PPT and $GR(\rho) \leq \epsilon$, we save $\rho$ in our dataset as SEP;
	\item If $\rho$ is NPT and $GR(\rho) > \epsilon$, we save $\rho$ in our dataset as NPT;
	\item Encode $\rho$ in $\vec{c}$ using Eqs. (\ref{eq-complete-rho}) and (\ref{eq-linear-inversion}) choosing $SU(3)\otimes SU(3)$ generators.
\end{enumerate}

\subsection{Dataset with artificial PPTES}

As we know that the probability of drawing a PPTES is low, another possible strategy is to increase the odds using what we call \emph{artificial PPTES} via edge states \cite{lewenstein2001}.

In this methodology, we compute \emph{Optimal Decomposable Witnesses} (d-OEW) \cite{lewenstein2000,Brandao2005} for the GR, obtaining the edge state $\xi^{PPT}$, lying in the frontier between PPT and NPT sets (\emph{cf.} Fig \ref{fig:state-space}). We say this is the minimal amount of mixing with any state required to become a PPT state. 

Cases where $\xi^{SEP} \neq \xi^{PPT}$ are particularly useful, since we can obtain a PPTES with any convex combination between them \cite{lewenstein2001} (\emph{cf.} Fig \ref{fig:state-space}). We say they are different in trace-norm given a tolerance of $10^{-3}$.

Thus, we can summarize the algorithm to generate our dataset as follows:

\begin{enumerate}
	\item Draw a random density matrix $\rho$ according to the Hilbert-Schmidt measure;
	\item Compute $GR(\rho)$ using an $\epsilon$-OEW and obtain its edge state $\xi^{SEP}$;
	\item Check whether $\rho$ is a PPT state;
	\item If $\rho$ is PPT and $GR(\rho) > \epsilon$, we save $\rho$ in our dataset as PPTES;
	\item If $\rho$ is PPT and $GR(\rho) \leq \epsilon$, we save $\rho$ in our dataset as SEP;
	\item If $\rho$ is NPT and, consequently, $GR(\rho) > \epsilon$, we save $\rho$ in our dataset as NPT. 
	\item If $\rho$ is NPT, we also quantify $GR(\rho)$ computing an (d-OEW) and its edge state $\xi^{PPT}$;
	\item Check whether $\xi^{SEP}$ and $\xi^{PPT}$ are different in trace norm, \emph{e.g}, $\frac{1}{2}\norm{\xi^{SEP} -\xi^{PPT} }_1 > 10^{-3}$;
	\item If they are different, we have $\delta = \frac{1}{2}\xi^{PPT} + \frac{1}{2}\xi^{SEP}$  and save it as PPTES.
	\item Compute $GR(\delta)$ using an $\epsilon$-OEW;
	\item If $GR(\rho) > \epsilon$ we save $\delta$ in our dataset as PPTES;
\end{enumerate}

Since we are including artificially created PPTES in addition to randomly drawn ones, we need to assess whether we have a fair sample in this subset. In Tab. \ref{tab:avg-fidelity}, we have the average fidelities, 

\begin{equation}
F(\rho,\sigma) = \left[ \tr\left(\sqrt{\sqrt{\rho} \sigma\sqrt{\rho}}\right) \right]^2,    
\end{equation}

over states in our dataset. Note a similar average fidelity of PPTES comparing with separable states and, as expected, NPT states are more distant since they have a larger volume in Hilbert space \cite{zyczkowski1998}. 

\begin{table}
\centering
\caption{Average fidelity $\langle F \rangle$ between states in the sample.}
\label{tab:avg-fidelity}
\begin{tabular}{cccc}
\hline \hline
&Class & $\langle F \rangle$ & \\
\hline
&Sample & 0.7879&\\
&SEP    & 0.7955&\\
&PPTES  & 0.7965& \\
&NPT    & 0.7712&\\
\hline \hline
\end{tabular}

\end{table}

The final dataset consists of 3254 states of each class in the triple (SEP/PPTES/NPT), with half random and half artificial PPTES, totaling 9762 states. Although we could benefit from a larger dataset, it took about a month using a cluster of 18 cores and 90GB of RAM to sample ten thousand random states and calculate their $\epsilon$-OEW within the desired numerical precision. 

\section{\label{sec:automl} Automated Machine Learning} 

Automated Machine Learning (autoML) stands for the automation of some or even all of the following tasks in the ML workflow: feature engineering; model tuning; ensembling; and model deployment. Open source projects such as TPOT \cite{le2019scaling}, auto-sklearn \cite{NIPS2015_5872}, AutoKeras \cite{jin2019auto}, and Auto-PyTorch \cite{mendoza-automlbook18a} are common tools used in this endeavor. It is not an exaggeration saying that autoML is a form of learning about learning \cite{Guyon2019}. 

We have a threefold motivation to use autoML: (i) we are dealing with a proved NP-HARD problem without a huge dataset; (ii) autoML can provide useful insights about specific tasks, such as different pre-processing schemes; (iii) autoML has a good cost-benefit in terms of implementation time and performance. 

(i) is the crux of the matter, so automatically trying different learning algorithms and architectures is paramount to achieve optimum performance \cite{high-bias}. (ii) although autoML was designed for non-experts, an experienced modeller can extract important insights when analyzing partial results, such as the need for \emph{Principal Component Analysis} (PCA) \cite{wold1987}, feature importance, and so on. Finally, (iii) was achieved with good accuracy and precision within a few hours of training and a couple of minutes to implement it.

Here, we used supervised learning methods, in which we know the output for each input in the sample, with the goal of inferring a function that best maps inputs to outputs. Common supervised learning tasks are: (i) classification, when one needs to map input to discrete output (labels), or (ii) regression, when one seeks to map input to a continuous output set. Typical supervised learning algorithms include: Logistic Regression (LR) \cite{seber2003}, Naive Bayes (NB) \cite{schutze2008}, Support Vector Machines (SVM) \cite{platt1998,cristianini2000}, K-Nearest Neighbour (KNN) \cite{keller1985}, Decision Tree (DT) \cite{breiman1984}, Extreme Gradient Boosting (XGB) \cite{chen2015}, Random Forest (RF) \cite{archer2008}, and Artificial Neural Networks (ANN) \cite{Anderson1992}, to cite a few. Our autoML schemes received data points from features $\mathbf{X}$ with corresponding target vector $\mathbf{y}$, and modeled the conditional probability $p(\mathbf{y} | \mathbf{X})$, resulting in a general rule mapping $\mathbf{X} \to \mathbf{y}$. 

In the next section, we have results for three different tasks: (i) binary classification of PPT states of two qutrits as SEP or entangled with positive PPTES; (ii) multi-classification of SEP vs PPTES vs NPT; and (iii) a regression learning of the Generalized Robustness (GR), an entanglement quantifier defined in Eq. (\ref{eq:gr}). Although (iii) seams a completely different problem, we can see a regression task as a classification using a discretized continuous target \cite{regascla}. So, we can use this regression to estimate the amount of entanglement in the quantum state and validate our classifiers.

\section{\label{sec:results} Results and Discussion}

\subsection{\label{sub:feat} Feature Importance}

In the previous section, we mentioned the possibility of extracting valuable insights when analyzing partial results using autoML. For example, we can use TPOT \cite{le2019scaling} for feature importance analysis. TPOT runs a genetic program to explore different pipelines given the raw input and output. It automatically runs standard tests, \emph{e.g.}, analysis of data type, $z$-score, check the relevance of higher powers of the inputs, dimensionality reduction via PCA, hyper-parameters optimization, and so on. It is a good entry point for any autoML project, since we can investigate if one feature is more relevant than the others, which is crucial to diminish the number of used features in a learning process, reducing the computational effort and increasing the robustness against over-fitting \cite{high-bias}.

We tested the feature importance of our dataset using TPOT with $10$ generations and a population size of $30$. The optimization took about $6$ hours to conclude using a standard PC and we confirmed that all $80$ features contribute equally to solve the problem. Therefore, we cannot reduce the dimensionality of the data without losing considerable information. Since our features $c_i$ (from Eq. (\ref{eq-linear-inversion}) represents the inner product of $\rho$ against linearly independent observables ($SU(3) \otimes SU(3)$ generators), this result is expected.

This stress the difference when using random states to train machines capable of discerning between unknown coefficients. In case we want to analyze a family of states composed of fewer parameters, the problem simplifies enormously and it is possible to achieve excellent results \cite{Gao2018,Yang2019}.

For the classification tests, in addition to the features of the vector $\vec{c} \, \in \, \mathbb{R}^{80}$, we also included its exponentials, $\exp(c_{i})\, \forall \, i=1,\dots ,80$ and the second-order terms, $c_i\cdot c_j \, \forall \, i=1,\dots,80 \text{ and } j=i,\dots, 80$, totalling $3400$ features. All features were standardized to z-scores, \emph{i.e.}, centered to have zero mean and scaled to have standard deviation of one.

\subsection{\label{sub:binary} Binary Classification (SEP \emph{vs.} PPTES)}

\begin{table}[t!]
\centering
\caption{Testing performance for different ML techniques using the same dataset for the binary classification SEP \emph{vs.} PPTES. We show the best performing one in boldface.  The autoML techniques are TPOT, auto-sklearn, AutoKeras and Auto-PyTorch. The conventional methods are ensemble method AdaBoost, artificial neural network (ANN), support vector machine (SVM), K-nearest neighbors (KNN), decision tree (DT), and random forest (RF).}
\label{table:bin-clas}
\begin{tabular}{ccccc}
\hline \hline
 ML tecnhique & Test (\%) &  PPTES (\%) & SEP (\%) \\ 
 \hline
  \textbf{TPOT} &  \textbf{75.3} &  \textbf{76.6} &  \textbf{74.1}\\
 auto-sklearn & 73.4  & 71.9 & 75.0 \\
 AutoKeras & 41.6  & 40.0 & 36.0 \\
 Auto-PyTorch & 62.2 & 63.7  & 60.8  \\
 AdaBoost & 66.9 & 65.7 & 68.1  \\
 ANN & 58.4  & 57.0 & 59.9\\
 SVM & 70.5  & 67.1 & 73.7 \\
 KNN & 48.4 & 0.2 & 99.8 \\
 DT & 53.0 & 49.6 & 56.5 \\
 RF  & 70.3  & 73.2 &  67.4\\
 \hline \hline
\end{tabular}
\end{table}

After deciding the features, the first task is training binary classifiers to discern between separable and entangled states with positive partial transposition. This is especially important when the Peres-Horodecki criterion fails and we need to decide whether $\rho$ is entangled.

To assess the overall performance of each classifier, we split the dataset in two parts: $\backsim 80\% $ for training and $\backsim 20\% $ for testing.  The idea is to test a trained model with ``new information'', so we can ensure the model's robustness against over-fitting. Thus, we say we have a good model when it achieves a high performance in the testing phase.

\begin{figure}[htpb]
\includegraphics[width=\columnwidth]{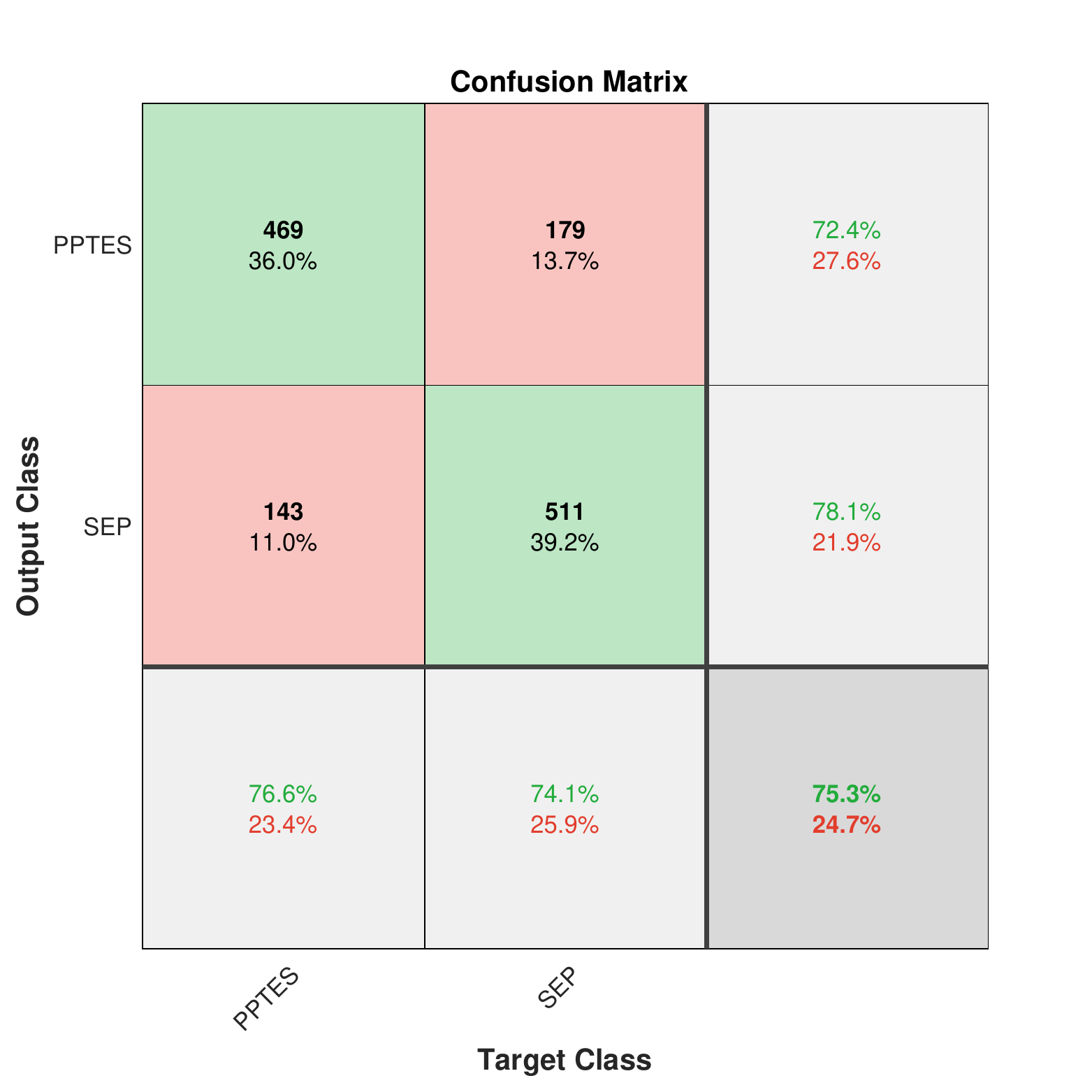}
\caption{\label{fig:confusion2} The testing Confusion Matrix of TPOT classifier. The rows correspond to the output classes (SEP/PPTES) and the columns to the true classes (also SEP/PPTES), having the correctly classified inputs in the diagonal entries and the incorrectly classified in the off-diagonal, with both number of states and percentage of the total number of observations. The last row and column show the percentages of all states predicted to belong to each class that are (in)correctly classified. TPOT achieved an overall performance of $75.3\%$. }
\end{figure}

We trained different models using different ML techniques within a time window of $6$ hours in a standard PC. Their testing performances are in Tab. \ref{table:bin-clas}, with the best model in boldface. Notice that all autoML strategies (TPOT, auto-sklearn, AutoKeras, Auto-PyTorch, and AdaBoost) outperformed standard algorithms (ANN, SVM, KNN, DT, and RF). In the testing step, TPOT classified SEP with $ 74.1\%$ accuracy and PPTES with  $ 76.6\%$, with a overall performance of  $75.3\%$. The \emph{Confusion Matrix} (CM) \cite{aaron} for this test is in Fig. \ref{fig:confusion2}.

We faced over-fitting in all models due to the curse of dimensionality and the limited amount of data \cite{high-bias}. In some cases, we could get close to $100\%$ accuracy in the training step, but with terrible performance in the testing phase. Nonetheless, the accuracy of about $75.3\%$ for out-of-sample data using TPOT indicates that ML can be a potential new tool to investigate bound entanglement in high-dimensional systems.

\subsection{\label{sub:multi-clas} Multi-classification (SEP \emph{vs.} PPTES \emph{vs.} NPT)}

Although we can check whether a quantum state is PPT/NPT efficiently, after the successful classification in the previous section, we want to investigate one more class in our autoML framework for a complete entanglement description of two qutrits. The task at hand is a trinary classification of separable, entangled states with positive partial transposition and entangled states with negative partial transposition. 

\begin{figure}[htpb]
\includegraphics[width=\columnwidth]{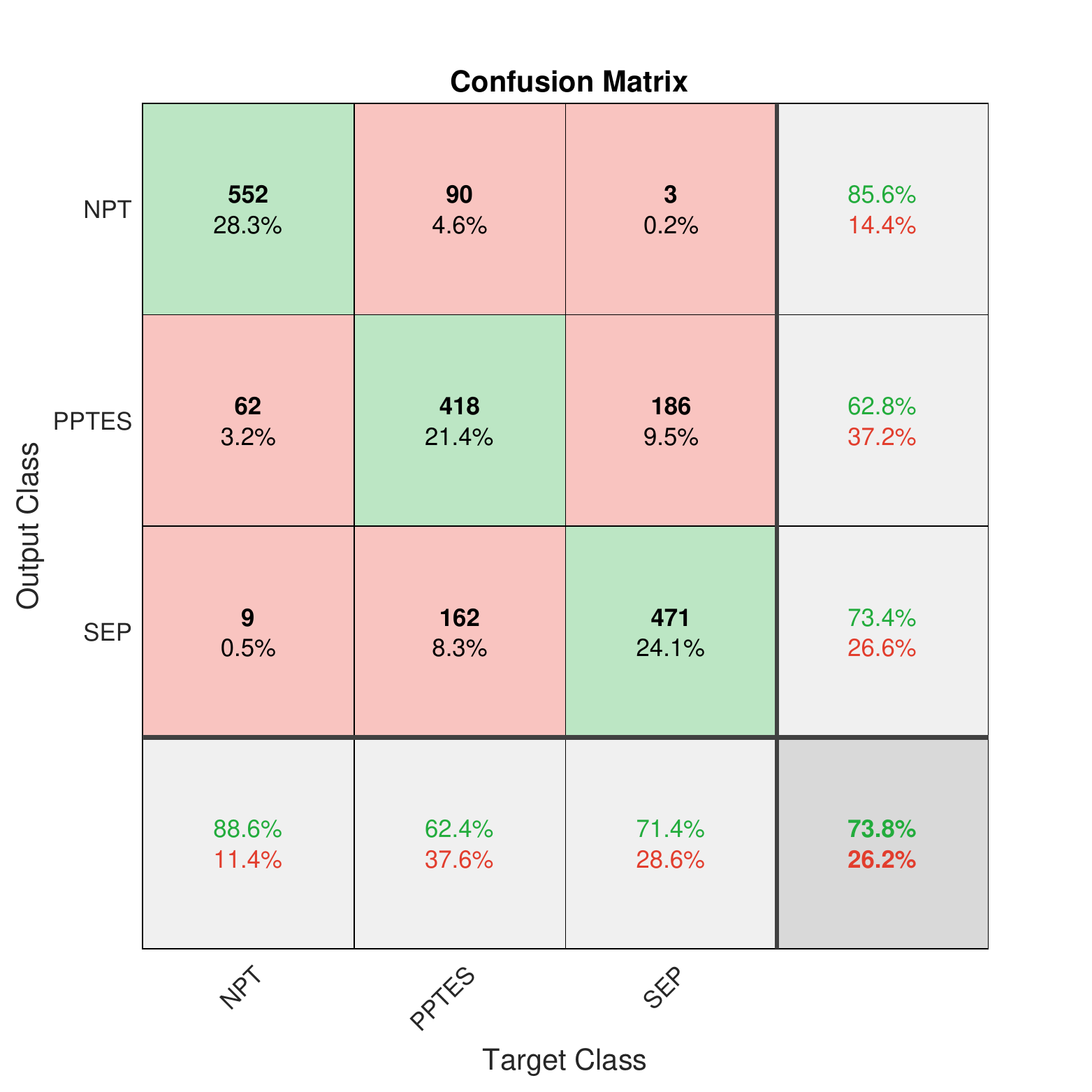}
\caption{\label{fig:confusion-multi} The testing Confusion Matrix of TPOT classifier. The rows correspond to the output classes (SEP/PPTES/NPT) and the columns to the true classes (also SEP/PPTES/NPT), having the correctly classified inputs in the diagonal entries and the incorrectly classified in the off-diagonal, with both number of states and percentage of the total number of observations. The last row and column show the percentages of all states predicted to belong to each class that are (in)correctly classified. TPOT achieved an overall performance of $73.8\%$.}
\end{figure}

For this task, we split the dataset in $\backsim 80\% $ for training and $\backsim 20\% $ for testing, and run the same algorithms from the previous section within a time window of $6$ hours in a standard PC. We present the results in Tab. \ref{table:multi-clas} with the best model in boldface. Notice that TPOT still outperforms other strategies, accurately classifying SEP with $71.4 \%$,  PPTES $62.4\%$, and NPT with $88.6\%$, with a overall performance of $73.8\%$. Its confusion matrix for the testing phase is in Fig. \ref{fig:confusion-multi}.

\begin{table}[htpb]
\centering
\caption{\label{table:multi-clas} Different machine learning techniques applied to the dataset and the corresponding accuracy for the multi-classification. We show the best performing one in boldface. The autoML techniques are TPOT, auto-sklearn, AutoKeras, and Auto-PyTorch. The ensemble method AdaBoost. The conventional methods are artificial neural network (ANN) support vector machine (SVM), K-nearest neighbors (KNN), decision tree (DT), random forest (RF).}
\begin{tabular}{ccccc}
\hline \hline
 ML tecnhique & Test (\%) & NPT (\%) & PPTES (\%) & SEP (\%) \\ 
 \hline
 \textbf{TPOT} & \textbf{73.8} & \textbf{88.6} & \textbf{62.4} & \textbf{71.4}\\
 auto-sklearn & 71.9 & 86.1 & 53.2 & 77.2 \\
 AutoKeras & 52.5 & 53.0 & 44.5 & 54.9 \\
 Auto-PyTorch  & 42.7 & 68.7 & 14.0 & 45.1 \\
 AdaBoost & 64.2 & 95.6 & 48.1 & 73.5\\
 ANN & 56.0 & 70.3 & 42.4 & 55.6\\
 SVM & 72.9 & 86.5 & 53.7 & 69.6 \\
 KNN & 34.1 & 0.4 & 0.6 & 99.6 \\
 DT & 43.1 & 48.0 & 35.1 & 45.9\\
 RF  & 69.2 & 85.0 & 45.6 & 76.4 \\
 \hline \hline
\end{tabular}
\end{table}

It is impressive how TPOT and auto-sklearn could deal with the curse of dimensionality, achieving an outstanding accuracy for this challenging scenario. The classification set has now several boundaries (\emph{cf.} Fig. \ref{fig:state-space}): between SEP and PPTES; PPTES and NPT; and between SEP and NPT \footnote{Werner states \cite{werner1989} is a one-parameter family from SEP to NPT without passing through PPTES.}. A reduction of the overall performance was expected, especially with a high-dimensional dataset, possessing $3400$ features, and $9762$ samples. We noticed a poor performance only in the deep learning (DL) autoML methods, known to need a large amount of instances to achieve a convergence \cite{aaron}. Notwithstanding, with overall accuracy of $73.8\%$ for out-of-sample data using TPOT indicates that ML can be a potential new tool to investigate bound entanglement in high-dimensional systems.

\subsection{\label{sub:reg} Regression Task}

As we mentioned in Sec. \ref{sec:automl}, regression is a harder problem, we can see it as a multi-classification with a infinite number of classes. Here, the target is the Generalized Robustness, an entanglement quantifier defined in Eq. (\ref{eq:gr}) and calculated for each sample in our dataset with an $\epsilon$-OEW, and $\epsilon = 10^{-5}$ (\emph{cf.} Sec. \ref{sec:dataset}).

In spite of starting with low-precision target values, and not possessing a huge dataset to fill the infinite classes, autoML still manages to recover a tendency line of the quantifier. The idea here is to use the regression to validate the previous classifiers and estimate the amount of entanglement in the state to some degree. Thus, higher values for the predicted GR can give us confidence about the entanglement of an unknown state. 

\begin{table}[htpb]
\centering
\caption{Different machine learning techniques applied to the dataset and their verified mean absolute error (MAE) on the test set. The autoML techniques are TPOT, AutoKeras, and Auto-PyTorch. The conventional methods are artificial neural network (ANN), support vector machine (SVM), K-nearest neighbors (KNN), decision tree (DT), random forest (RF).}
\label{table:reg}
\begin{tabular}{ccccc}
\hline \hline
 ML tecnhique & MAE \\ 
 \hline
 TPOT & 0.0373 \\
%  auto-sklearn & -$^1$ \\
 \textbf{AutoKeras} & \textbf{0.0335} \\
 Auto-PyTorch  & 0.0347 \\
 ANN & 0.0350 \\
 SVM & 0.0383 \\
 KNN & 0.0500 \\
 DT  & 0.0649 \\ 
 RF  & 0.0463  \\
 \hline \hline
\end{tabular}
\end{table}

In Tab. \ref{table:reg} we have a comparison of different ML techniques for the regression task, with the best model in boldface. Since Auto-sklearn still very incipient for this task, we chose to remove it from this comparison. The autoML also proved to be efficient in this task, outperforming standard algorithms. To achieve the best performance,  AutoKeras had an extra fine-tuning, called \emph{final fit}, which took an extra hour in addition to the $6$ hours in our time window. This extra fine-tune was responsible for a reduction of almost $5\%$ in the \emph{Mean Absolute Error} (MAE).

\begin{figure}[t!]
\includegraphics[width=\columnwidth]{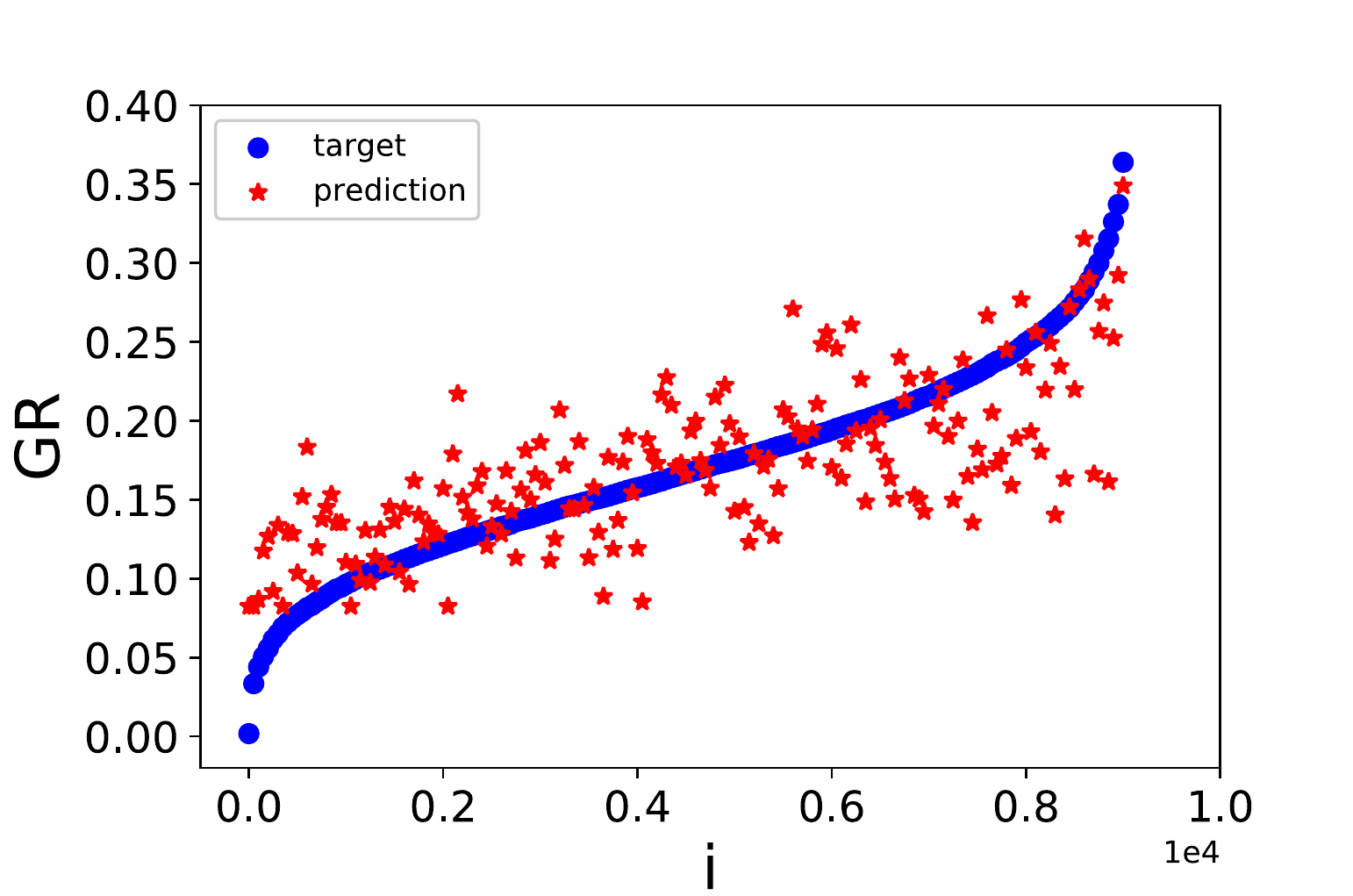}
\caption{\label{fig:reg_apyto} In blue (circle) the Generalized Robustness (\emph{cf.} Eq. (\ref{eq:gr})) and in red (star) the ML prediction considering around $9000$ test set points after an autoML model has been trained and deployed with AutoKeras package, with a mean absolute error (MAE) of $0.0335$.}
\end{figure}

We show the target values (sorted from minimum to maximum) and its corresponding regression using AutoKeras in Fig. \ref{fig:reg_apyto}, with a MAE of $L1 = 0.0335$. For a better plotting experience, we reduced the density of points by a factor of a hundred.

In case we want a model with a more accurate description of the quantifier, one should increase the number of samples in the dataset, so the density of points in each of these infinite classes represents the set accordingly.

Finally, we can establish a threshold in which we can trust the predictions as a function of our achieved MAE. For example, we can define a conservative threshold, $\delta \equiv 3 \times L1 \approx 0.1$, so, when the machine outputs GR greater than $\delta$ we say the state is entangled. Note that, for smaller values ($\text{GR} < 0.1$), the output is almost an upper-bound, meaning that the true value is smaller than the prediction. On the other hand, for large values ($\text{GR} > 0.25$) the predictions establish a lower-bond and the true value is larger than the prediction. For the intermediate region ($ 0.1 < \text{GR} < 0.25$), although we cannot assess upper/lower bounds, it satisfies the criterion to be larger than $\delta = 0.1$, and the state can be said to be entangled. 

\section{\label{sec:conclusion} Conclusion}

In this work, we used Automated Machine Learning (autoML) to successfully classify random quantum states of two qutrits as separable (SEP), PPT entangled state (PPTES), or NPT. We chose a scenario in which we possess enough data to reconstruct $\rho$ via linear system inversion, but no direct measurements of its entanglement. In addition to classification in terms of SEP/PPTES/NPT, we could also estimate the amount of entanglement in the state using autoML for a regression of the Generalized Robustness of Entanglement. In conclusion, if we use both classifiers and the regression, we can decide whether an unknown state is entangled and validate one test using others. 

We could successfully apply our framework even in the hardest instance of the problem, when the Peres-Horodecki criterion cannot decide whether a quantum state is separable or entangled with positive partial transposition. With autoML, for the binary (SEP \emph{vs.} PPTES) task, TPOT accurately classified SEP with $ 74.1\%$ and PPTES with  $ 76.6\%$, with an overall performance of  $75.3\%$. For the multi-classification (SEP \emph{vs.} PPTES \emph{vs.} NPT), TPOT accurately classified $71.4 \%$ of SEP,  $62.4\%$ of PPTES and $88.6\%$, of NPT states, with an overall performance of $ 73.8\%$. Finally, we achieved the mean absolute error of $0.0335$ for the regression problem with a conservative confidence threshold of $GR_{ML} > 0.1$ for entanglement assessment and validation of our classifiers.

Our result is less accurate than the one in \cite {Lu2018}, however we must emphasize we used no features intimately related with entanglement, that is, our machine is able to infer whether a two-qutrit system is entangled solely with tomograms. This opens the possibility of using the machine in real applications, since it eliminates the high computational cost involving any type of entanglement estimate.

To achieve models with higher performance in any of the presented tests, one should increase the number of samples. We trained our models with a dataset containing both random and artificial PPTES states (\emph{cf.} Sec \ref{sec:dataset}) and we have a strong numerical evidence that artificially created PPTES using the method in Sec. \ref{sec:dataset} is a good strategy to overcome the low probability to draw a random PPTES state using the Hilbert-Schmidt measure. 

This computational effort can pay off when we can decide the entanglement of a random state efficiently and with high probability. So, a global repository of correctly labeled states could help the community to train better machines and also validate the results, since we would have a larger number of groups working with the same dataset. This could be paramount in real-time applications, like quantum communication protocols.

We can also conclude that, given the hardness of the problem, we have a strong evidence that autoML can be beneficial to supervised ML tasks of all sizes and scales, including problems with smaller datasets, paving the way for future investigations using this framework.

% \textcolor{red}{Another important remark in this work is we demonstrate that in small or even intermediate complexity scale supervised machine learning tasks in science one should give automated machine learning at least a benchmark position, as it was the most successful approach in all of our examples, paving the way for future investigations in this direction.}

\begin{acknowledgements}
We acknowledge the financial support by Brazilian agencies CAPES, CNPq, and INCT-IQ (National Institute of Science and Technology for Quantum Information). We also thank SeTIC-UFSC for the computational time in its cluster. AC acknowledges UFAL for a paid license for scientific cooperation at UFRN and the John Templeton Foundation via the Grant Q-CAUSAL No. 61084, the Serrapilheira Institute (Grant No. Serra-1708-15763) and CNPQ (Grant No. $423713/2016-7$).
\end{acknowledgements}


\begin{thebibliography}{10}

\bibitem{shor1994}
Peter~W Shor.
\newblock Algorithms for quantum computation: Discrete logarithms and
  factoring.
\newblock In {\em Proceedings 35th annual symposium on foundations of computer
  science}, pages 124--134. Ieee, 1994.

\bibitem{vandersypen2001}
Lieven~MK Vandersypen, Matthias Steffen, Gregory Breyta, Costantino~S Yannoni,
  Mark~H Sherwood, and Isaac~L Chuang.
\newblock Experimental realization of shor's quantum factoring algorithm using
  nuclear magnetic resonance.
\newblock {\em Nature}, 414(6866):883, 2001.

\bibitem{Ursin2007}
R.~Ursin, F.~Tiefenbacher, T.~Schmitt-Manderbach, H.~Weier, T.~Scheidl,
  M.~Lindenthal, B.~Blauensteiner, T.~Jennewein, J.~Perdigues, P.~Trojek,
  B.~\"{O}mer, M.~F\"{u}rst, M.~Meyenburg, J.~Rarity, Z.~Sodnik, C.~Barbieri,
  H.~Weinfurter, and A.~Zeilinger.
\newblock Entanglement-based quantum communication over
  144{\hspace{0.167em}}km.
\newblock {\em Nature Physics}, 3(7):481--486, June 2007.

\bibitem{Vaziri2002}
Alipasha Vaziri, Gregor Weihs, and Anton Zeilinger.
\newblock Experimental two-photon, three-dimensional entanglement for quantum
  communication.
\newblock {\em Physical Review Letters}, 89(24), November 2002.

\bibitem{Grblacher2006}
Simon Gr\"{o}blacher, Thomas Jennewein, Alipasha Vaziri, Gregor Weihs, and
  Anton Zeilinger.
\newblock Experimental quantum cryptography with qutrits.
\newblock {\em New Journal of Physics}, 8(5):75--75, May 2006.

\bibitem{Peres1996}
Asher Peres.
\newblock Separability criterion for density matrices.
\newblock {\em Physical Review Letters}, 77(8):1413--1415, August 1996.

\bibitem{Horodecki1996}
Micha{\l} Horodecki, Pawe{\l} Horodecki, and Ryszard Horodecki.
\newblock Separability of mixed states: necessary and sufficient conditions.
\newblock {\em Physics Letters A}, 223(1-2):1--8, November 1996.

\bibitem{Horodecki1998}
Micha{\l} Horodecki, Pawe{\l} Horodecki, and Ryszard Horodecki.
\newblock Mixed-state entanglement and distillation: Is there a
  {\textquotedblleft}bound{\textquotedblright} entanglement in nature?
\newblock {\em Physical Review Letters}, 80(24):5239--5242, June 1998.

\bibitem{Horodecki1999}
Pawe{\l} Horodecki, Micha{\l} Horodecki, and Ryszard Horodecki.
\newblock Bound entanglement can be activated.
\newblock {\em Physical Review Letters}, 82(5):1056--1059, February 1999.

\bibitem{Bennett1999}
Charles~H. Bennett, David~P. DiVincenzo, Tal Mor, Peter~W. Shor, John~A.
  Smolin, and Barbara~M. Terhal.
\newblock Unextendible product bases and bound entanglement.
\newblock {\em Physical Review Letters}, 82(26):5385--5388, June 1999.

\bibitem{Breuer2006}
Heinz-Peter Breuer.
\newblock Optimal entanglement criterion for mixed quantum states.
\newblock {\em Physical Review Letters}, 97(8), August 2006.

\bibitem{Sents2016}
Gael Sent{\'{\i}}s, Christopher Eltschka, and Jens Siewert.
\newblock Quantitative bound entanglement in two-qutrit states.
\newblock {\em Physical Review A}, 94(2), August 2016.

\bibitem{Shor2003}
Peter~W. Shor, John~A. Smolin, and Ashish~V. Thapliyal.
\newblock Superactivation of bound entanglement.
\newblock {\em Physical Review Letters}, 90(10), March 2003.

\bibitem{Gurvits2003}
Leonid Gurvits.
\newblock {Classical complexity and quantum entanglement}.
\newblock In {\em Journal of Computer and System Sciences}, volume~69, pages
  448--484, mar 2004.

\bibitem{Gurvits2003a}
Leonid Gurvits.
\newblock Classical deterministic complexity of edmonds' problem and quantum
  entanglement.
\newblock In {\em Proceedings of the thirty-fifth annual ACM symposium on
  Theory of computing}, pages 10--19. ACM, 2003.

\bibitem{horodecki2009}
Ryszard Horodecki, Pawe{\l} Horodecki, Micha{\l} Horodecki, and Karol
  Horodecki.
\newblock Quantum entanglement.
\newblock {\em Reviews of modern physics}, 81(2):865, 2009.

\bibitem{guhne2002}
O~G{\"u}hne, P~Hyllus, D~Bru{\ss}, A~Ekert, M~Lewenstein, C~Macchiavello, and
  A~Sanpera.
\newblock Detection of entanglement with few local measurements.
\newblock {\em Physical Review A}, 66(6):062305, 2002.

\bibitem{cavalcanti2006}
Daniel Cavalcanti and Marcelo~O Terra~Cunha.
\newblock Estimating entanglement of unknown states.
\newblock {\em Applied physics letters}, 89(8):084102, 2006.

\bibitem{maciel2009}
Thiago~O Maciel and Reinaldo~O Vianna.
\newblock Viable entanglement detection of unknown mixed states in low
  dimensions.
\newblock {\em Physical Review A}, 80(3):032325, 2009.

\bibitem{lima2010}
G~Lima, ES~G{\'o}mez, A~Vargas, RO~Vianna, and C~Saavedra.
\newblock Fast entanglement detection for unknown states of two spatial
  qutrits.
\newblock {\em Physical Review A}, 82(1):012302, 2010.

\bibitem{Dunjko2018}
Vedran Dunjko and Hans~J Briegel.
\newblock Machine learning {\&} artificial intelligence in the quantum domain:
  a review of recent progress.
\newblock {\em Reports on Progress in Physics}, 81(7):074001, June 2018.

\bibitem{Kieferov2017}
M{\'{a}}ria Kieferov{\'{a}} and Nathan Wiebe.
\newblock Tomography and generative training with quantum boltzmann machines.
\newblock {\em Physical Review A}, 96(6), December 2017.

\bibitem{Torlai2018}
Giacomo Torlai, Guglielmo Mazzola, Juan Carrasquilla, Matthias Troyer, Roger
  Melko, and Giuseppe Carleo.
\newblock Neural-network quantum state tomography.
\newblock {\em Nature Physics}, 14(5):447--450, February 2018.

\bibitem{Torlai2017}
Giacomo Torlai and Roger~G. Melko.
\newblock Neural decoder for topological codes.
\newblock {\em Physical Review Letters}, 119(3), July 2017.

\bibitem{Krastanov2017}
Stefan Krastanov and Liang Jiang.
\newblock Deep neural network probabilistic decoder for stabilizer codes.
\newblock {\em Scientific Reports}, 7(1), September 2017.

\bibitem{August2017}
Moritz August and Xiaotong Ni.
\newblock Using recurrent neural networks to optimize dynamical decoupling for
  quantum memory.
\newblock {\em Physical Review A}, 95(1), January 2017.

\bibitem{Gao2018}
Jun Gao, Lu-Feng Qiao, Zhi-Qiang Jiao, Yue-Chi Ma, Cheng-Qiu Hu, Ruo-Jing Ren,
  Ai-Lin Yang, Hao Tang, Man-Hong Yung, and Xian-Min Jin.
\newblock Experimental machine learning of quantum states.
\newblock {\em Physical Review Letters}, 120(24), June 2018.

\bibitem{Yang2019}
Mu~Yang, Chang liang Ren, Yue chi Ma, Ya~Xiao, Xiang-Jun Ye, Lu-Lu Song,
  Jin-Shi Xu, Man-Hong Yung, Chuan-Feng Li, and Guang-Can Guo.
\newblock Experimental simultaneous learning of multiple nonclassical
  correlations.
\newblock {\em Physical Review Letters}, 123(19), November 2019.

\bibitem{Canabarro2019}
Askery Canabarro, Samura{\'{\i}} Brito, and Rafael Chaves.
\newblock Machine learning nonlocal correlations.
\newblock {\em Physical Review Letters}, 122(20), May 2019.

\bibitem{Lu2018}
Sirui Lu, Shilin Huang, Keren Li, Jun Li, Jianxin Chen, Dawei Lu, Zhengfeng Ji,
  Yi~Shen, Duanlu Zhou, and Bei Zeng.
\newblock Separability-entanglement classifier via machine learning.
\newblock {\em Physical Review A}, 98(1), July 2018.

\bibitem{Canabarro2019b}
Askery Canabarro, Felipe~Fernandes Fanchini, Andr\'e~Luiz Malvezzi, Rodrigo
  Pereira, and Rafael Chaves.
\newblock Unveiling phase transitions with machine learning.
\newblock {\em Phys. Rev. B}, 100:045129, Jul 2019.

\bibitem{aaron}
Ian Goodfellow, Yoshua Bengio, and Aaron Courville.
\newblock {\em Deep Learning}.
\newblock MIT Press, 2016.
\newblock \url{http://www.deeplearningbook.org}.

\bibitem{nielsen2000}
M.~Nielsen and I.~Chuang.
\newblock {\em Quantum Computation and Quantum Information}.
\newblock Cambridge University Press, 1 edition, 2000.

\bibitem{vidal1999}
Guifr\'e Vidal and Rolf Tarrach.
\newblock Robustness of entanglement.
\newblock {\em Phys. Rev. A}, 59:141--155, Jan 1999.

\bibitem{le2019scaling}
TT~Le, W~Fu, and JH~Moore.
\newblock Scaling tree-based automated machine learning to biomedical big data
  with a feature set selector.
\newblock {\em Bioinformatics (Oxford, England)}, 2019.

\bibitem{NIPS2015_5872}
Matthias Feurer, Aaron Klein, Katharina Eggensperger, Jost Springenberg, Manuel
  Blum, and Frank Hutter.
\newblock Efficient and robust automated machine learning.
\newblock In C.~Cortes, N.~D. Lawrence, D.~D. Lee, M.~Sugiyama, and R.~Garnett,
  editors, {\em Advances in Neural Information Processing Systems 28}, pages
  2962--2970. Curran Associates, Inc., 2015.

\bibitem{jin2019auto}
Haifeng Jin, Qingquan Song, and Xia Hu.
\newblock Auto-keras: An efficient neural architecture search system.
\newblock In {\em Proceedings of the 25th ACM SIGKDD International Conference
  on Knowledge Discovery \& Data Mining}, pages 1946--1956. ACM, 2019.

\bibitem{mendoza-automlbook18a}
Hector Mendoza, Aaron Klein, Matthias Feurer, Jost~Tobias Springenberg,
  Matthias Urban, Michael Burkart, Max Dippel, Marius Lindauer, and Frank
  Hutter.
\newblock Towards automatically-tuned deep neural networks.
\newblock In Frank Hutter, Lars Kotthoff, and Joaquin Vanschoren, editors, {\em
  AutoML: Methods, Sytems, Challenges}, chapter~7, pages 141--156. Springer,
  December 2018.
\newblock To appear.

\bibitem{steiner2003}
Michael Steiner.
\newblock Generalized robustness of entanglement.
\newblock {\em Phys. Rev. A}, 67(5):054305, 2003.

\bibitem{may2010}
Robert~J May, Holger~R Maier, and Graeme~C Dandy.
\newblock Data splitting for artificial neural networks using som-based
  stratified sampling.
\newblock {\em Neural Networks}, 23(2):283--294, 2010.

\bibitem{zyczkowski2011}
Karol {\.Z}yczkowski, Karol~A Penson, Ion Nechita, and Benoit Collins.
\newblock Generating random density matrices.
\newblock {\em Journal of Mathematical Physics}, 52(6):062201, 2011.

\bibitem{zyczkowski1998}
Karol {\.Z}yczkowski, Pawe{\l} Horodecki, Anna Sanpera, and Maciej Lewenstein.
\newblock Volume of the set of separable states.
\newblock {\em Physical Review A}, 58(2):883, 1998.

\bibitem{zyczkowski2003}
Karol Zyczkowski and Hans-J{\"u}rgen Sommers.
\newblock Hilbert--schmidt volume of the set of mixed quantum states.
\newblock {\em Journal of Physics A: Mathematical and General}, 36(39):10115,
  2003.

\bibitem{sommers2004}
Hans-J{\"u}rgen Sommers and Karol {\.Z}yczkowski.
\newblock Statistical properties of random density matrices.
\newblock {\em Journal of Physics A: Mathematical and General}, 37(35):8457,
  2004.

\bibitem{braunstein1996}
Samuel~L Braunstein.
\newblock Geometry of quantum inference.
\newblock {\em Physics Letters A}, 219(3-4):169--174, 1996.

\bibitem{zyczkowski2001}
Karol Zyczkowski and Hans-J{\"u}rgen Sommers.
\newblock Induced measures in the space of mixed quantum states.
\newblock {\em Journal of Physics A: Mathematical and General}, 34(35):7111,
  2001.

\bibitem{Horodecki2001}
Micha{\l} Horodecki, Pawe{\l} Horodecki, and Ryszard Horodecki.
\newblock Separability of n-particle mixed states: necessary and sufficient
  conditions in terms of linear maps.
\newblock {\em Physics Letters A}, 283(1-2):1--7, 2001.

\bibitem{brandao2004a}
Fernando~GSL Brandao and Reinaldo~O Vianna.
\newblock Separable multipartite mixed states: operational asymptotically
  necessary and sufficient conditions.
\newblock {\em Physical review letters}, 93(22):220503, 2004.

\bibitem{Brandao2005}
Fernando G. S.~L. Brand\~ao.
\newblock Quantifying entanglement with witness operators.
\newblock {\em Phys. Rev. A}, 72:022310, Aug 2005.

\bibitem{doherty2004}
Andrew~C. Doherty, Pablo~A. Parrilo, and Federico~M. Spedalieri.
\newblock Complete family of separability criteria.
\newblock {\em Phys. Rev. A}, 69:022308, Feb 2004.

\bibitem{lewenstein2001}
Maciej Lewenstein, B~Kraus, P~Horodecki, and JI~Cirac.
\newblock Characterization of separable states and entanglement witnesses.
\newblock {\em Physical Review A}, 63(4):044304, 2001.

\bibitem{lewenstein2000}
Maciej Lewenstein, B~Kraus, JI~Cirac, and P~Horodecki.
\newblock Optimization of entanglement witnesses.
\newblock {\em Physical Review A}, 62(5):052310, 2000.

\bibitem{Guyon2019}
Isabelle Guyon, Lisheng Sun-Hosoya, Marc Boull{\'e}, Hugo~Jair Escalante,
  Sergio Escalera, Zhengying Liu, Damir Jajetic, Bisakha Ray, Mehreen Saeed,
  Mich{\`e}le Sebag, Alexander Statnikov, Wei-Wei Tu, and Evelyne Viegas.
\newblock {\em Analysis of the AutoML Challenge Series 2015--2018}, pages
  177--219.
\newblock Springer International Publishing, Cham, 2019.

\bibitem{high-bias}
Pankaj Mehta, Marin Bukov, Ching-Hao Wang, Alexandre~G.R. Day, Clint
  Richardson, Charles~K. Fisher, and David~J. Schwab.
\newblock A high-bias, low-variance introduction to machine learning for
  physicists.
\newblock {\em Physics Reports}, 810:1–124, May 2019.

\bibitem{wold1987}
Svante Wold, Kim Esbensen, and Paul Geladi.
\newblock Principal component analysis.
\newblock {\em Chemometrics and intelligent laboratory systems}, 2(1-3):37--52,
  1987.

\bibitem{seber2003}
George~AF Seber and Christopher~John Wild.
\newblock Nonlinear regression. hoboken.
\newblock {\em New Jersey: John Wiley \& Sons}, 62:63, 2003.

\bibitem{schutze2008}
Hinrich Sch{\"u}tze, Christopher~D Manning, and Prabhakar Raghavan.
\newblock {\em Introduction to information retrieval}, volume~39.
\newblock Cambridge University Press Cambridge, 2008.

\bibitem{platt1998}
John Platt.
\newblock Sequential minimal optimization: A fast algorithm for training
  support vector machines.
\newblock Technical Report MSR-TR-98-14, April 1998.

\bibitem{cristianini2000}
Nello Cristianini, John Shawe-Taylor, et~al.
\newblock {\em An introduction to support vector machines and other
  kernel-based learning methods}.
\newblock Cambridge university press, 2000.

\bibitem{keller1985}
James~M Keller, Michael~R Gray, and James~A Givens.
\newblock A fuzzy k-nearest neighbor algorithm.
\newblock {\em IEEE transactions on systems, man, and cybernetics},
  (4):580--585, 1985.

\bibitem{breiman1984}
L~Breiman and JH~Friedman.
\newblock Ra olshen and cj stone,“.
\newblock {\em Classification and regression trees}, 1984.

\bibitem{chen2015}
Tianqi Chen, Tong He, Michael Benesty, Vadim Khotilovich, and Yuan Tang.
\newblock Xgboost: extreme gradient boosting.
\newblock {\em R package version 0.4-2}, pages 1--4, 2015.

\bibitem{archer2008}
Kellie~J Archer and Ryan~V Kimes.
\newblock Empirical characterization of random forest variable importance
  measures.
\newblock {\em Computational Statistics \& Data Analysis}, 52(4):2249--2260,
  2008.

\bibitem{Anderson1992}
Dave Anderson and George McNeill.
\newblock Artificial neural networks technology.
\newblock {\em Kaman Sciences Corporation}, 258(6):1--83, 1992.

\bibitem{regascla}
Raied Salman and Vojislav Kecman.
\newblock Regression as classification.
\newblock In {\em Conference Proceedings - IEEE SOUTHEASTCON}, pages 1--6, 03
  2012.

\bibitem{werner1989}
Reinhard~F Werner.
\newblock Quantum states with einstein-podolsky-rosen correlations admitting a
  hidden-variable model.
\newblock {\em Physical Review A}, 40(8):4277, 1989.

\end{thebibliography}
\end{document}